\documentclass{elsart}

 \usepackage{epsfig}

\usepackage{amssymb}
\usepackage{amsmath}

\begin{document}

\begin{frontmatter}

\title{Dynamics of the quantum Duffing oscillator in the driving
induced bistable regime 
\thanksref{birthday}}
\thanks[birthday]{This work is dedicated to Prof.\ Phil Pechukas.}

\author{V.\ Peano  and }
\author{M.\ Thorwart\corauthref{cor}}
\corauth[cor]{Corresponding author}
\ead{thorwart@thphy.uni-duesseldorf.de}

\address{Institut f\"ur Theoretische Physik IV, 
Heinrich-Heine-Universit\"at D\"usseldorf, Universit\"atsstr.\ 1, 
40225 D\"usseldorf, Germany}

\begin{abstract}
% Text of abstract
We investigate the nonlinear response of an anharmonic monostable quantum mechanical 
resonator to strong external periodic driving. The driving thereby induces an effective 
bistability in which resonant tunneling can be identified. Within the framework of a 
Floquet analysis,   an effective Floquet-Born-Markovian master equation with 
time-independent coefficients   
can be established which can be solved straightforwardly. Various effects including 
resonant tunneling and multi-photon transitions will be described. Our model finds 
applications in nano-electromechanical devices such as vibrating suspended nano-wires 
as well as in non-destructive read-out procedures for superconducting quantum bits 
involving the nonlinear response of the read-out SQUID. 
\end{abstract}
\begin{keyword}
Quantum dissipation \sep tunneling \sep driven systems \sep nonlinear response 
\PACS  03.65.-w \sep 62.25.+g \sep 62.30.+d \sep 03.65.Xp
\end{keyword}
\end{frontmatter}

% main text
\section{Introduction}
\label{intro}
Classical nonlinear systems subjected to strong periodic external driving often have 
several stable stationary states for which the amplitudes and phases of the
forced vibrations differ in size \cite{Nayfeh,Holmes,Jackson}. 
One of the simplest theoretical models
which show the coexistence of two stable states induced by external driving 
is the well-known classical Duffing
oscillator. An anharmonic statically monostable potential can be driven into a
dynamically bistable regime showing various interesting features of
non-linear response \cite{Nayfeh,Holmes,Jackson}, 
such as hysteresis, period doubling, and thermal
activation 
when finite temperatures are considered. The external driving
field with frequency $\omega$ induces an effective dynamic bistability 
which  is manifest  by the non-monotonous dependence of
the amplitude $A$ of the stationary vibrations for varying $\omega$. For
the classical system where all potential energies are allowed, this
response curve $A(\omega)$
is smooth showing only two points of bifurcation for the related
bistability. If the control parameter $\omega$ is additionally varied
adiabatically, hysteretical jumps between the two stable states occur. 
If additional thermal noise is added to the system, the regime of
bistability shrinks due to thermal escape of the metastable state. 

The main subject of this work is to investigate the corresponding driven 
{\em quantum
mechanical\/} system. The presence of time-dependent driving typically 
adds several interesting features to  the properties of 
the time-independent quantum system, see for instance Refs.\   \cite{Pechukas76,DrivenSys}.  
In this work, 
the focus is laid on the non-linear response of the
driven quantum mechanical anharmonic oscillator in the presence of 
an Ohmic heat bath. We show that the nonlinear response curve $A(\omega)$
displays beyond its characteristic shape additional quantum mechanical 
resonances which are related to multi-photon absorptions. Additionally, we
show that there exists a separation of time-scales indicating that the
stationary state is reached by  quantum tunneling from the dynamically
induced metastable to the globally stable state. In fact, by tuning the
control parameter $\omega$,  several resonant tunneling transitions can be
identified as resonances in the corresponding tunneling rate. 
 
The paper is organized in the following way: In Section \ref{model}, the
system Hamiltonian is introduced. The latter is periodic in time, which
allows the application of Floquet theory.  Since we are interested in the stationary
state in presence of (weak) dissipation, we introduce also a set of harmonic
oscillators representing an Ohmic heat bath. An efficient way to determine
the dynamics of the system is the use of a Born-Markovian master equation in
the Floquet picture (Section \ref{masterequation}) yielding a simple master
equation with time-independent rate coefficients. After straight forward
diagonalization, the stationary oscillation amplitude $A$ and the phase
$\varphi$ follow. In
Sections \ref{amplitude} and \ref{phase}, the nonlinear response of $A$ and
$\varphi$ depending on the various parameters is studied in detail. 
In Section \ref{restun}, resonant tunneling is investigated. Finally, 
 in Section \ref{discussion}, we
discuss the applicability of our model to experimental systems 
before we conclude. 

%%%%%%%%%%%%%%%
\section{The quantum Duffing oscillator}
\label{model}
The Hamiltonian for the driven anharmonic oscillator has the form 
\begin{equation}
H_S(t)=\frac{p^2}{2m}+\frac{m \omega_0^2}{2}x^2 + \frac{\alpha}{4}x^4+ x f \cos(\omega
t) \, . \label{ham}
\end{equation}
Here, $m$ and $\omega_0$ are the mass and the harmonic frequency of the resonator, respectively, while $\alpha$
gives the strength of the nonlinearity. We focus on the case $\alpha>0$ of hard non-linearities, 
where the undriven potential is
monostable. The external driving is characterized by the amplitude 
$f$ and the frequency $\omega$. As it will become clear below, the driving induces an effective bistability in
which quantum tunneling can be identified.  

 We include the effect of the environment by a bath of harmonic oscillators 
 which are bilinearly coupled to the system  
 with the coupling constants $c_j$ \cite{Weiss99}. 
The  Hamiltonian for the bath and the coupling to the system is given by its standard form 
\begin{equation}
H_B=\frac{1}{2}\sum_j \frac{p_j^2}{m_j} + m_j \omega_j^2 \left( x_j- \frac{c_j}{m_j \omega_j^2}x\right)^2. %\, .
\end{equation}
We focus on the generic case of an Ohmic bath with the spectral density 
\begin{equation}
J(\omega)=\frac{\pi}{2}\sum_j\frac{c_j^2}{m_j\omega_j}\delta(\omega-\omega_j)=
m\gamma\omega e^{-\omega/\omega_c}\, , 
\end{equation}
with  damping constant $\gamma$ and cut-off 
frequency $\omega_c$.  The total Hamiltonian is $H(t)=H_S(t)+H_B$. 

To proceed, we scale $H(t)$ to dimensionless quantities 
such that the energies are in units of $\hbar \omega_0$ while the lengths are scaled in units of 
$x_0\equiv \sqrt{\frac{\hbar}{m \omega_0}}$. Put differently, we
formally set $m=\hbar=\omega_0=1$. 
The nonlinearity parameter $\alpha$ is scaled in units of 
$\alpha_0\equiv \hbar \omega_0/x_0^4$, while the driving amplitude is given in units of $f_0\equiv 
\hbar \omega_0 / x_0$. Moreover, we scale temperature in units of $T_0\equiv \hbar \omega_0 / k_B$ while the
damping strengths are measured with respect to $\omega_0$. 

To investigate the dynamical behavior of the driven resonator, it is convenient
to use the periodicity of $H_S(t)$ with respect to time and switch to the Floquet picture  \cite{Kohler97}, 
the later being
equivalent to a transformation to  the rotating frame.  The Floquet or quasi-energies $\varepsilon_\alpha$ 
follow from the solution of the eigenvalue equation 
\begin{equation}
\left[ H_S(t) -i\hbar \frac{\partial}{\partial t} \right]
|\phi_\alpha (t) \rangle = \varepsilon_\alpha |\phi_\alpha (t) \rangle \, ,
\end{equation}
with the Floquet states $|\phi_\alpha (t) \rangle$. 
The quasienergies $\varepsilon_\alpha$ are defined up to a multiple integer 
of $\hbar \omega$, i.e., the state $|\phi^{(n)}_{\alpha}(t)\rangle 
= e^{i n \omega t}|\phi_{\alpha}(t)\rangle$ 
is also an eigenstate of the Floquet Hamiltonian, but with the eigenvalue 
$\varepsilon_{\alpha,n}=\varepsilon_\alpha + n \hbar \omega$. 
This feature prevents us from a global ordering of the quasienergies which, however, can be achieved 
with  the mean energies obtained after averaging over one driving period, i.e.,  
\begin{equation}
\overline{E}_\alpha
=\sum_n (\varepsilon_{\alpha}+n \hbar \omega) \langle c_{\alpha,n} | 
c_{\alpha,n}\rangle\, ,
\end{equation}
with the Fourier components of the Floquet states \cite{Kohler97} 
\begin{equation}
|  c_{\alpha,n}\rangle = \frac{\omega}{2 \pi} \int_0^{2\pi/\omega} dt \, e^{in
\omega t } |\phi_{\alpha}(t)\rangle \, .
\end{equation}
%. 
%%%%%%%%%%%%%%%%%
\section{Dynamics of the quantum Duffing oscillator}
\label{masterequation}
\subsection{Floquet-Born-Markovian master equation}
The dynamics of the resonator in the regime of weak coupling to the bath can be efficiently 
described by a Born-Markovian master equation in the Floquet picture \cite{Kohler97} 
for the elements  $\rho_{\alpha\beta} (t)\equiv \langle \phi_{\alpha}(t) | \rho|\phi_{\beta}(t)\rangle$ 
of the reduced density operator $\rho$ after the harmonic bath has been traced out in the usual way. 
For weak damping, the dissipative influence of the bath is relevant only on a time scale much larger 
than the driving period $T_\omega=2 \pi / \omega$. Thus, the time-dependent coefficients which are periodic 
in time with period $T_\omega$ can safely be replaced by their average over one driving period 
({\em moderate rotating wave approximation\/} \cite{Kohler97}). This yields a simplified master equation 
with time-independent coefficients which reads
\begin{equation}
\dot{\rho}_{\alpha\beta} (t) = \sum_{\alpha' \beta'} 
\mathcal{M}_{\alpha\beta,\alpha' \beta'} \rho_{\alpha' \beta'} (t) \,  
\label{floqme}
\end{equation}
with 
\begin{equation}
\mathcal{M}_{\alpha\beta,\alpha' \beta'} = -\frac{i}{\hbar} 
\left(
\varepsilon_\alpha - \varepsilon_\beta
\right) \delta_{\alpha\alpha'} \delta_{\beta\beta'} + 
\mathcal{L}_{\alpha\beta,\alpha' \beta'} \, . 
\label{supop}
\end{equation}
The first term on the r.h.s.\ describes the coherent time evolution of the pure system while the second term
contains the  transition rates describing the influence of the dissipative bath. It reads \cite{Kohler97}
\begin{eqnarray}
\mathcal{L}_{\alpha\beta,\alpha' \beta'} & = &  \sum_n \left( N_{\alpha \alpha', n} +
N_{\beta \beta', n} \right) X_{\alpha \alpha', n} X_{\beta' \beta,-n}   - \delta_{\beta \beta'} 
\sum_{\beta'', n}  N_{\beta'' \alpha', n}  X_{\alpha \beta'', -n} X_{\beta''
\alpha', n}  \nonumber \\ 
& & - \delta_{\alpha \alpha'} 
\sum_{\alpha'', n}  N_{\alpha'' \beta', n}  X_{ \beta' \alpha'', -n} X_{\alpha''
\beta, n} \, . 
\label{ratecoeff}
\end{eqnarray}
with  the coefficients 
\begin{eqnarray}
N_{\alpha \beta, n} &=& N(\varepsilon_\alpha-\varepsilon_\beta + n \hbar \omega)\,
, \mbox{\hspace{4ex}} N(\varepsilon) = \frac{m \gamma \varepsilon}{\hbar^2} 
\frac{1}{e^{\varepsilon/k_B T}-1} \, , \nonumber \\
X_{\alpha \beta, n} &=& \frac{\omega}{2 \pi} 
\int_0^{2\pi/\omega} dt \, e^{-in\omega t} \langle \phi_{\alpha}(t)|x|\phi_{\beta}(t)\rangle
\, .
\end{eqnarray}
The operator $\mathcal{M}$ in Eq.\  (\ref{supop}) is a $M^2\times M^2-$matrix, when the total Hilbert space of the
anharmonic resonator has been truncated to the $M-$dimensional subspace. For practical purposes, we set $M=12$
throughout this work. Note that we have confirmed convergence with respect to $M$ 
for all results shown below. With that, $\mathcal{M}$ 
can be readily diagonalized numerically by standard means. This can be 
formalized in terms of the diagonalization transformation $S$ 
by the eigenvalue equation
\begin{equation}
\sum_{\mu \nu, \mu' \nu'} ( S^{-1})_{\alpha \beta, \mu \nu} \mathcal{M}_{\mu \nu, \mu' \nu'} 
S_{\mu' \nu', \alpha' \beta'} = \Lambda_{\alpha \beta} \delta_{\alpha, \alpha'} \delta_{\beta, \beta'} \, .
\end{equation}
Here, $\Lambda_{\alpha \beta}$ denote the eigenvalues of the operator $\mathcal{M}$. 
Since the master equation (\ref{floqme}) conserves the trace of
$\rho$, there is an eigenvalue $\Lambda^{\infty}_{\alpha \beta}=0$ which
characterizes the stationary solution $\rho_{\alpha \beta}^{\infty}$. 
The remaining eigenvalues all have a 
negative real part leading to a decay of the corresponding mode with time. 
 Due to the structure of the master equation (\ref{floqme}), there exist two
classes of eigenvalues: (i) the first class having an imaginary part of
zero consists of individual eigenvalues (associated to {\em relaxation\/}), 
and (ii) the second class having
non-zero imaginary parts (associated to 
{\em dephasing\/}) consists of pairs of  complex
conjugated eigenvalues.  
The eigenvalues can be ordered according to the size of the absolute
value of their real parts.
As it turns out, there is one pair of complex conjugated eigenvalues,
i.e., $\Lambda_{\alpha\beta, {\rm min}}^{(1)}=\Lambda_{\alpha\beta,
{\rm min}}^{(2)*}$ with the smallest non-zero absolute 
value of the real part. Moreover, we find that that pair of 
eigenvalues is clearly separated in size
from the remaining ones with respect to the real parts indicating a
separation of time scales. 
That pair of eigenvalues is responsible for the tunneling dynamics at
long times, as will be discussed in
Section \ref{restun}. 
Finally, the solution of the master equation  (\ref{floqme}) can be 
formally written as 
\begin{equation}
\rho_{\alpha \beta}(t) = \rho_{\alpha \beta}^{\infty} + \sum_{\mu \nu, \mu' \nu'} 
S_{\alpha \beta, \mu \nu}  (S^{-1})_{\mu \nu, \mu' \nu'} e^{\Lambda_{\mu \nu}t} \rho_{\mu'\nu'}(t=0) \, .
\label{solution}
\end{equation}
For convenience, we have chosen the initial time $t=0$. 
\subsection{Observable: Stationary oscillation amplitude}
We are interested in calculating the asymptotic expectation value $\langle x \rangle_{t,{\rm as}}$ 
of the position operator. This is the
quantity which can directly be compared to its classical counterpart being the solution 
of the classical Duffing equation. It reads 
\begin{equation}
\langle x \rangle_{t,{\rm as}} = \sum_n c_n e^{in \omega t} \, 
\label{xampl}
\end{equation}
with the coefficients 
\begin{equation}
c_n = A_n e^{i\varphi_n}=\sum_{\alpha \beta} \rho_{\alpha \beta}^\infty X_{\beta \alpha,n} \, . 
\label{xcoeff}
\end{equation}
Since the oscillator is driven by a cosine-shaped external force, it will oscillate in the asymptotic limit 
also with a phase-shifted cosine if the excitation frequency $\omega$ remains close to $\omega_0$. 
Higher harmonics could be generated in this nonlinear system but they are not in the focus of our interest in
this work. Thus, only the terms $n=\pm 1$ contribute in the Fourier expansion and we obtain 
\begin{equation}
\langle x \rangle_{t,{\rm as}} =  A \cos (\omega t+\varphi) \, , \mbox{\hspace{3ex}} \omega \approx \omega_0 
\end{equation}
with the amplitude $A\equiv 2 A_1$ and the phase $\varphi \equiv \varphi_1$ of the first harmonic 
of the Fourier expansion. These two quantities are used to study the nonlinear response of the 
anharmonic resonator in the stationary long-time limit. The short time dynamics of such a type of 
master equation  
is an interesting issue by itself since it is related to the question of complete positivity
 and to the question of factorizing initial conditions \cite{Pechukas94,Ankerhold02}. 
 Moreover, we note that the master equation (\ref{floqme}) is valid only in the case of weak system-bath
 coupling. For the opposite limit of strong coupling (quantum Smoluchowski limit), 
 different techniques \cite{Thorwart97,Thorwart00,Ankerhold01a,Ankerhold01b,Machura04} 
 have to be applied. 
\subsection{Classical Duffing oscillator}
The corresponding classical oscillator (at zero temperature) is 
the well-known Duffing oscillator \cite{Nayfeh,Holmes,Jackson}. It shows a rich variety of features including
regular and chaotic motion. In this work, we focus on the parameter regime where only regular motion
occurs. The nonlinear
response of its amplitude $A$ can be calculated perturbatively \cite{Nayfeh}. One obtains 
the response $A(\omega)$ as the solution of the equation
\begin{equation}
\omega - \omega_0 = \frac{3}{8} \frac{\alpha}{m\omega_0}A^2 \pm \left(\frac{f^2}{4 m^2 \omega_0^2 A^2} -
\frac{\gamma^2}{4}\right)^{1/2} \, .
\label{ampl}
\end{equation}
Its characteristic form is shown in the inset of Fig.\  \ref{fig.1}. 
For weak driving strengths, the response as a function of the driving frequency $\omega$ 
has the well-known form of the harmonic oscillator with the maximum at $\omega=\omega_0$. 
For increasing
driving strength, the resonance grows and bends away from the $\omega=\omega_0$-axis 
towards larger frequencies (since $\alpha>0$). The locus of the maximal amplitudes is given by the 
parabola \cite{Nayfeh} $\omega-\omega_0=\frac{3}{8} \frac{\alpha}{m\omega_0}A^2$, 
which is often called the {\em backbone curve\/}. 
Most importantly, a bistability develops with two adjacent stable branches and one intermediate 
unstable branch. This bistability is connected with a hysteretical jump phenomenon which can be 
probed if the
driving frequency $\omega$ is adiabatically increased or decreased. The hysteresis is 
maximal for zero
temperature. At finite temperature, it is reduced since the particle can escape from 
the metastable local minimum
to the adjacent global minimum via thermal hopping \cite{Pechukas81,PGH,HanggiRMP} 
before the deterministic switching 
point is reached \cite{Dykman79,Datta01}. Note that  for stronger
driving amplitudes, also bifurcations and
period doublings can occur \cite{Nayfeh,Holmes,Jackson} 
which we do not address in the present work. 

The nonlinear response of the phase $\varphi$ can be determined perturbatively in a similar way. One obtains
\cite{Nayfeh} 
\begin{equation}
\varphi = - \arctan  \frac{\gamma A}{2((\omega-\omega_0)A - \frac{3}{8}\frac{\alpha A^3}{\omega_0})},  
\end{equation}
where $A$ is the solution of Eq.\  (\ref{ampl}). The curve $\varphi(\omega)$ also has two stable
branches with an unstable intermediate branch and displays similar hysteretical jump phenomena as 
the amplitude response. 

As we will show in the following, the nonlinear response is qualitatively
different in the corresponding  quantum system. The discrete quasi-energy spectrum allows for multi-photon
excitations which yield discrete resonances in the amplitude response profile. Moreover, the dynamically
generated bistability allows for an escape of the system out of the metastable state  via 
resonant quantum tunneling. This generates characteristic resonances in the tunneling rate when the external
frequency $\omega$, which plays the role of a control parameter, is tuned. 
%%%%%%%%%%%%%%%%%%%%%%%
\section{Amplitude response}
\label{amplitude}
\begin{figure}[t]
\begin{center}
\epsfig{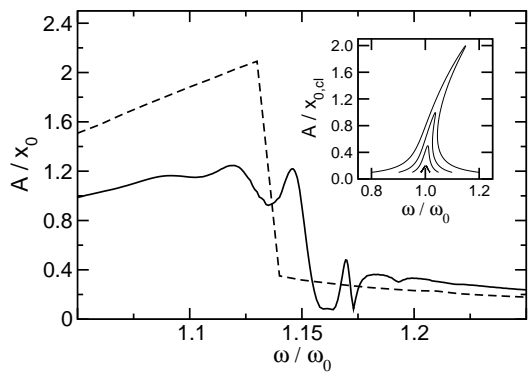}
\caption{Amplitude $A$ of the expectation value $\langle x (t)\rangle$ 
in units of $x_0=\sqrt{\frac{\hbar}{m\omega_0}}$ 
for varying the driving frequency $\omega$. 
Parameters are $k_B T=0.1 \hbar \omega_0, 
\alpha=0.1\alpha_0, f=0.1 f_0, \gamma = 0.005 \omega_0$. 
Dashed line: Results of the classical Duffing oscillator at 
$T=0$ with the remaining parameters being the same. 
%for the same parameters. 
Inset: Amplitude $A$ of the classical Duffing oscillator for varying 
driving frequencies. The driving strength $f$ increases from bottom 
to top. \label{fig.1}\vspace*{20mm}}
\end{center}
\end{figure}

A typical response profile for the amplitude $A(\omega)$ is shown in Fig.\ 
\ref{fig.1}. The  shoulder-like shape is a remnant of the classical form of
the response which is indicated by the dashed line. In the quantum case, clear
resonances can be observed at particular values of the driving frequency. 

\begin{figure}[t]
\begin{center}
\epsfig{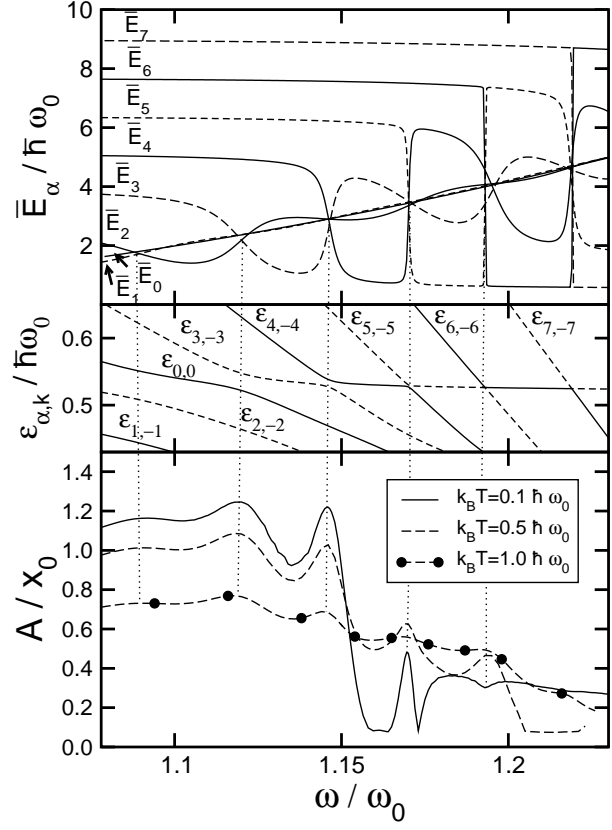}
\caption{Average energies  $\overline{E}_\alpha$ (top), 
quasienergy levels $\varepsilon_{\alpha, k}$ (middle) and 
amplitude $A$ of the fundamental mode for 
 varying driving frequencies $\omega$.  
The remaining parameters are $\alpha=0.1\alpha_0, f=0.1 f_0$ 
and $\gamma=0.005 \omega_0$. \label{fig.2}\vspace*{20mm}}
\end{center}
\end{figure}

The resonances can be understood as discrete multi-photon transitions
occurring when the $N$-th multiple of the field quantum $\hbar \omega$ 
equals the corresponding energy gap in the anharmonic oscillator. 
By inspection of the 
associated quasienergy spectrum shown in  Fig.\  \ref{fig.2}, one can see that the distinct
resonances occur at multiple degenerate 
avoided level crossings of the quasienergy level $\varepsilon_N$ 
of the $N$-th Floquet state with the quasienergy level $\varepsilon_0$ of the Floquet 
groundstate. In physical terms, multi-photon excitations occur in the resonator and the corresponding
populated Floquet state dominates then the position expectation value $\langle x \rangle$ with its
large value of the amplitude, see Eqs.\  (\ref{xampl},\ref{xcoeff}). 

The width of the $N$-th resonance is determined by the minimal splitting at
the avoided quasienergy level crossing which
is equal to the $N$-photon Rabi frequency. The latter can be evaluated perturbatively for 
weak driving amplitudes \cite{Dykman04,Larsen76} upon applying a
rotating-wave approximation (RWA). For the resulting Hamiltonian, 
one finds that to lowest order in $f/(\omega -
\omega_N)$, the $N$-photon Rabi frequency at $\omega_N$ decreases
exponentially with $N$ implying that the resonances are sharper for larger $N$. 
For small frequencies, 
the broad peaks overlap strongly and lead to a shoulder-like profile which 
 is similar to the classical result (dashed line in Fig.\  \ref{fig.1}).
 Although the RWA yields a qualitatively correct picture, a quantitative
 comparison shows noticeable deviations from the results for the full
 anharmonic resonator and is not pursued further in this work. Note that a similar system has been 
 investigated in Ref.\  \cite{Vogel88} in the context of  a dispersive optical bistability. 
 There, a nonlinear Hamiltonian has been derived from the Duffing oscillator using a RWA,  
 similar to that used in Ref.\ \cite{Dykman88,Dykman04}. A rather involved 
 matrix continued-fraction method also revealed a bistability. 
 However, the numerical procedure to obtain those results is rather cumbersome. 
 In contrast, our approach is numerically straightforward since it only involves a simple numerical 
 diagonalization of a matrix.  
\subsection{Tuning system parameters}
Since the location of the multi-photon resonances is determined by the system Hamiltonian, it is
interesting to see how they depend on varying the system parameters $\alpha$ and $f$. 

\subsubsection{Varying the nonlinearity coefficient $\alpha$}

\begin{figure}[t]
\begin{center}
\epsfig{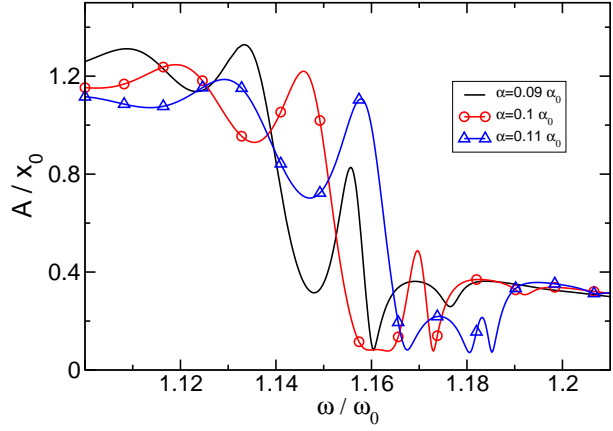}
\caption{(Color online) Response curve $A(\omega)$ 
for different values of the nonlinearity  $\alpha$. 
Remaining parameters are $k_B T=0.1 \hbar \omega_0, 
 f=0.1 f_0, \gamma = 0.005 \omega_0$. 
  \label{fig.3}\vspace*{20mm}}
\end{center}
\end{figure}

The spectrum of the system and therefore, the Floquet spectrum, depends
strongly on the nonlinearity coefficient $\alpha$. Increasing $\alpha$
increases the energy gaps between the succeeding eigenstates. Thus, the
multiphoton absorption processes occur at larger frequencies of the
driving field. This behavior is shown in Fig.\  \ref{fig.3}. The increase
of $\alpha$ leads in general to a shift of the response curve $A(\omega)$
towards larger frequencies $\omega$. Moreover, we observe that the height
of the $N$-photon peak decreases for increasing $\alpha$. This observation is
qualitatively in agreement with a perturbative treatment for weak driving 
within a rotating wave approximation \cite{Dykman04}.  

\subsubsection{Varying the driving amplitude $f$: Multiphoton antiresonance}

\begin{figure}[t]
\begin{center}
\epsfig{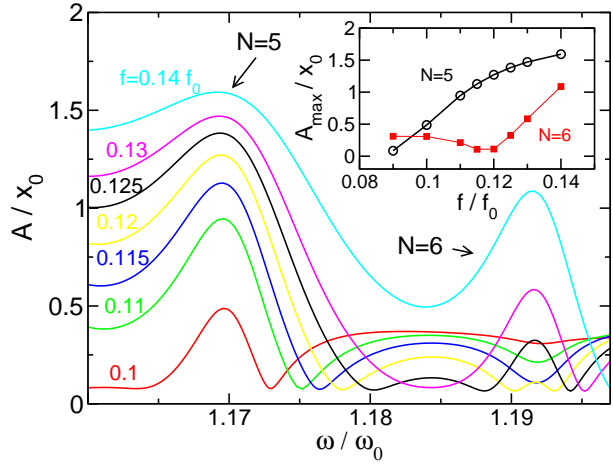}
\caption{(Color online) Five-and six-photon resonance of the response curve $A(\omega)$ 
for different values of the driving amplitude   $f$. Inset: Local extremum of the
$N-$photon resonance for varying driving strength. 
The six-photon resonance develops from an antiresonance for small driving
strengths to a true resonance for larger values of $f$. 
Remaining parameters are $k_B T=0.1 \hbar \omega_0, 
 \alpha=0.1\alpha_0, \gamma = 0.005 \omega_0$. 
  \label{fig.adriv}\vspace*{20mm}}
\end{center}
\end{figure}

The Floquet spectrum of the uncoupled system depends on the value of the
driving amplitude $f$. A perturbative analysis \cite{Dykman04} shows that
the $N$-photon Rabi frequency $\Omega_{R,N}$ which is given by the minimal splitting of
the quasienergy at the avoided quasienergy level crossing depends
crucially on $N$ and $f$. This, in turn, determines the behavior of the
$N$-photon resonance. The detailed results of the dependence on the $f$ are
shown in Fig.\  \ref{fig.adriv}. The $N=5$-photon peak grows for growing
$f$. Most interestingly, the $N=6$-photon peak displays a nontrivial
behavior. For weak driving, a $6$-photon antiresonance, i.e., a dip, 
develops. When the field amplitude is increased the antiresonance turns
into a true resonance which grows further for growing $f$. The particular
dependence of the local extrema are shown in the inset of Fig.\ 
\ref{fig.adriv}. 
As follows from Eqs.\  (\ref{xampl}) and (\ref{xcoeff}), the amplitude $A$ is 
determined by the sum $c_1=\sum_{\alpha \beta}\rho_{\alpha \beta}^{\infty} 
X_{ \beta\alpha,1}$. Thus, the product of the weights 
$\rho_{\alpha \beta}^{\infty}$ (which also contains the influence of the bath) and 
the matrix elements $X_{ \beta\alpha,1}$ (which is a property of the coherent 
driven system alone) determines the full shape of the amplitude. Notice that 
$\rho_{\alpha \beta}^{\infty}$ is in general not diagonal. This is different 
from the system considered in Ref.\  \cite{Dykman04} where only a coherent 
system without bath has been investigated. 

By varying the system parameters, we have modified implicitly both 
$\rho_{\alpha \beta}^{\infty}$ and $X_{ \beta\alpha,1}$. In the following section, 
only bath parameters and thus $\rho_{\alpha \beta}^{\infty}$ will be modified, 
leaving $X_{ \beta\alpha,1}$ unchanged.

%
%%%%%%%%%%%%%%%%%%%%%%%
\subsection{Dependence on bath parameters}
The crossover from antiresonant to resonant behavior occurs when an increased 
 driving amplitude $f$ increases the population of higher-lying Floquet states. 
In this subsection, we investigate the
role of the dissipative environment by tuning the bath temperature and the damping constant. 
\subsubsection{Varying temperature $T$}
\begin{figure}[t]
\begin{center}
\epsfig{figure=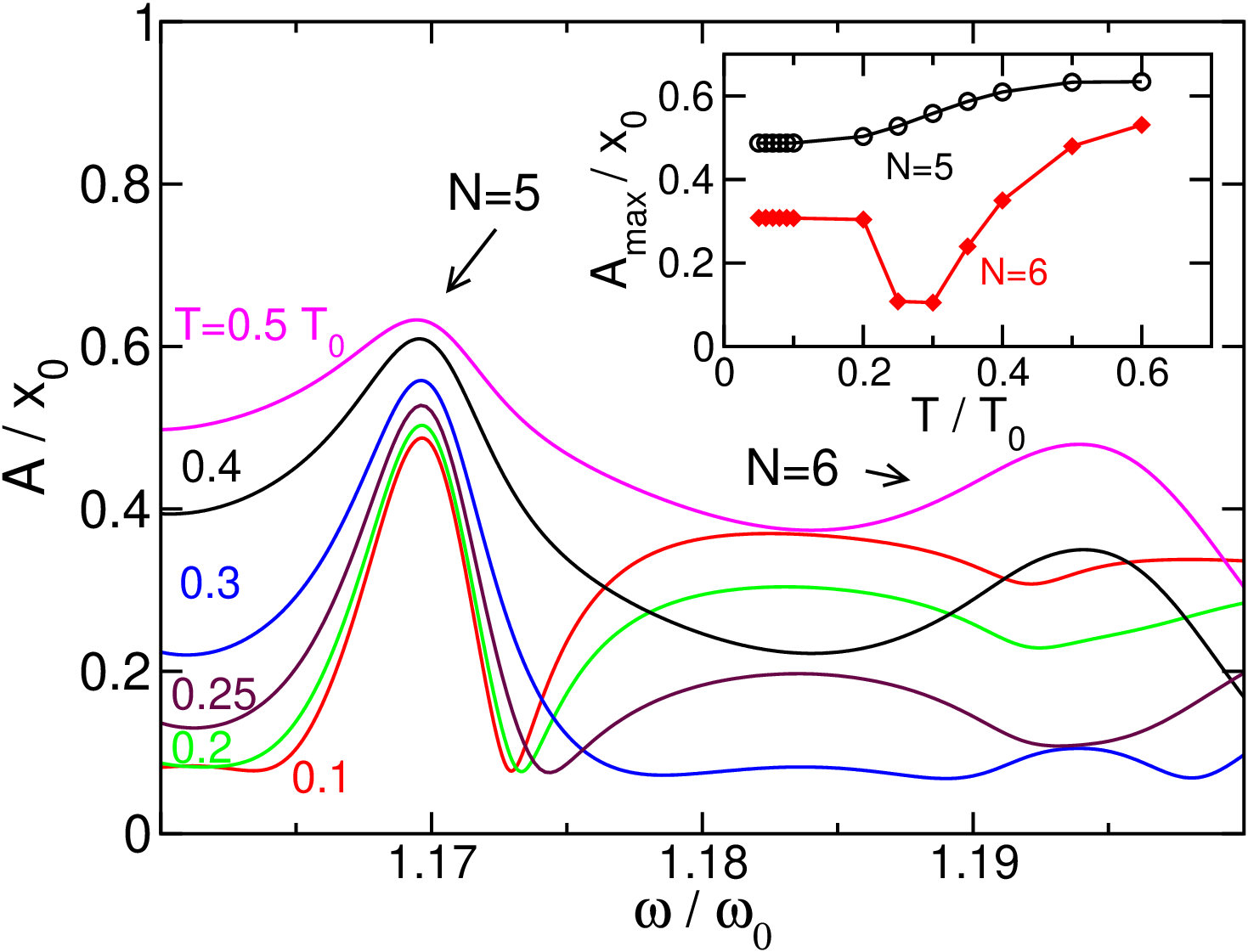,width=80mm,keepaspectratio=true}
\caption{(Color online) Five-and six-photon resonance of the response curve $A(\omega)$ 
for different values of the temperature   $T$. Inset: Local extremum of the
$N-$photon resonance for varying temperature. 
The six-photon resonance develops from an antiresonance for low $T$ 
 to a true resonance for higher $T$. 
Remaining parameters are $f=0.1 f_0, 
 \alpha=0.1\alpha_0, \gamma = 0.005 \omega_0$. 
  \label{fig.atemp}\vspace*{20mm}}
\end{center}
\end{figure}

Since increasing temperature $T$ leads also to an increased population of other Floquet modes, 
the transition from antiresonant to resonant behavior can also be expected to occur for growing $T$, 
at least in a certain temperature regime.  The result is shown in Fig.\ \ref{fig.atemp} for 
the 5-photon and 6-photon-resonance. The resonant peak for $N=5$ grows and broadens if temperature is
increased. In contrast, the antiresonant dip for $N=6$ shrinks for growing $T$ and turns into a 
resonant peak. This peculiar behavior can be interpreted as 
 thermally assisted cross-over from antiresonant to resonant behavior. 
The inset  in Fig.\ \ref{fig.atemp} shows the local extrema for $N=5$ and $N=6$. 
For even larger temperature $k_B T \approx \hbar \omega_0$, the characteristic peak structure is 
completely smeared out due to thermal broadening, see also Fig.\ \ref{fig.2}, lower panel. 
Then, the peaks overlap and the dynamic bistability is smeared out by thermal transitions between the two
(meta-)stable states. 
\subsubsection{Varying the damping constant $\gamma$}
The results for different damping constants $\gamma$ are shown in Fig.\  \ref{fig.adamp}. 
The five-photon resonance decreases when $\gamma$ is increased from $\gamma=0.001 \omega_0$ to 
$\gamma=0.01 \omega_0$. As also shown in the inset of Fig.\  \ref{fig.adamp}, the peak maximum decreases
monotonously for $N=5$. For the six-photon (anti-)resonance, we find a different behavior. For weak
damping, a sharp resonance occurs which is turned into an antiresonance for larger damping. The
non-monotonous dependence of the six-photon resonance is also shown in the inset of Fig.\  \ref{fig.adamp}. 
As it is the case for very large temperature (see above), 
the influence of a strong coupling to the bath gradually smears out and
finally destroys the resonances. 
%%%%%%%%%%%%%%%%%%%%%%%
\section{Phase response}
\label{phase}
Next, we address the nonlinear response of the phase $\varphi$. 
The corresponding classical Duffing oscillator shows an interesting nonlinear phase response 
$\varphi(\omega)$ including two stable branches \cite{Nayfeh}. 
The characteristic multi-photon transitions in the 
quantum version of the Duffing oscillator also shows up in the phase response profile. 
The results for different driving strengths $f$ are shown in Fig.\  \ref{fig.phidriv} 
for the regime where the five- and six-photon transition occur. For $\omega<\omega_0$ a phase 
shift $\varphi=-\pi$ is found. For $\omega\gg \omega_0$, the phase shift vanishes, $\varphi=0$. 
In the intermediate region $\omega \gtrsim \omega_0$, the multi-photon resonances induce also an 
antiresonance in the phase shift. The five-photon antiresonance is enhanced and broadened 
if the driving strength $f$ is increased from $f=0.09 f_0$ to $f=0.1 f_0$. 
Increasing the driving further to $f=0.11 f_0$ wipes out the antiresonance completely and the transition 
$\varphi=-\pi$ to $\varphi=0$ is shifted to larger values of $\omega$, where this development is repeated 
at the succeeding six-photon resonance. 

This behavior is also found when temperature is varied, see Fig.\  \ref{fig.phitemp}. Increasing $T$ leads to
a suppression of the transition of the phase shift $\varphi=-\pi \rightarrow \varphi=0$. In contrast,
increasing damping favors this transition at lower frequencies $\omega$, see Fig.\  \ref{fig.phidamp}. 

%%%%%%%%%%%%%%%%%%%
\section{Resonant tunneling in the driving induced bistability}
\label{restun}
The dynamic bistability of the steady state of the classical 
Duffing oscillator does not survive in the quantum system.  
The reason is that the system will escape the metastable state 
asymptotically 
via tunneling, similar to the case of the driven double-well potential 
\cite{Thorwart98}. Note also that, as a consequence, 
the hysteretical behavior is suppressed 
if the control parameter $\omega$ is varied truly adiabatically. 

Nevertheless, signatures of the dynamic bistability  
and tunneling can be found if we consider how the steady state is reached. 
For this, we show in  Fig.\  \ref{fig.timeresolv} the time evolution of 
the amplitude $A$ (local maxima of the vibrations) 
starting with the ground state of the 
undriven oscillator as the initial state. We observe fast 
oscillations at short times. They decay on a time scale 
$\gamma^{-1}$ which reflects ``intrawell'' relaxation 
in the metastable state.  Then, starting from a metastable 
state at intermediate times, a slow exponential 
decay towards the asymptotically globally stable state can be observed. 
This separation of time scales is a clear indication of tunneling from the 
meta- to the globally stable  in the dynamic bistability. 

The decay rate $\Gamma$ for this slow process (tunneling rate)  
is determined by the absolute value of the real part of the 
 eigenvalue $\Lambda_{\alpha\beta, {\rm min}}^{(1,2)}$ of the 
 operator $\mathcal{M}$ in Eq.\  (\ref{supop}), i.e., 
 $\Gamma= |{\rm Re}\Lambda_{\alpha\beta, {\rm min}}^{(1,2)}|$.  
 Results for the tunneling rate 
 as function of the control parameter $\omega$ are shown in Fig.\  \ref{fig.restun} 
 for two different damping constants $\gamma$.  
Most importantly, the tunneling rate shows resonances 
at the same values of the frequencies 
where the avoided crossings of the quasienergy levels occur 
(see dashed vertical lines). 
 The peaks in $\Gamma$ indicate resonant tunneling 
\cite{Thorwart98} 
from the meta- to the globally stable state 
both of which are dynamically induced. 
 Note the analogy to resonant tunneling
in a static double-well potential \cite{Thorwart98}. The role of the
eigenenergies in the static case is now played by the quasienergies
$\varepsilon_\alpha$ determining the coherent dynamics, see Eq.\  
(\ref{floqme}). In both cases, the avoided (quasi-)energy level crossings are
the origin of resonant tunneling. Nevertheless, the incoherent 
part of Eq.\  (\ref{floqme}) is crucial to
observe the resonant tunneling in this driving induced bistability. 

Furthermore, it is interesting to note that the resonant tunneling is 
enhanced if the coupling to the bath is increased from $\gamma=0.001 \omega_0$ to 
$\gamma=0.005 \omega_0$ (bath assisted resonant tunneling). We have also calculated the 
dependence of this phenomenon on temperature but we found a weak dependence in the interesting 
low-temperature regime (not shown).  
%
%%%%%%%%%%%%%%%%%%%%
\section{Discussion and conclusions}
\label{discussion}
The quantum Duffing oscillator is a  generic theoretical model which finds several applications 
in experimental systems. For instance, a suspended nanomechanical beam \cite{Roukes01,Roukes00} 
which is excited to transverse vibrations behaves as a damped anharmonic resonator. Several 
experimental groups have observed the behavior described by the {\em classical\/} 
Duffing oscillator \cite{Kroemmer00,Erbe00,Buks01,Husain03}. 
The transition to the quantum regime is currently in the focus of intense research 
\cite{Carr01,Werner04,Peano04,Ming03,Knobel03,LaHaye04}. Once, 
such kind of truly quantum-'mechanical' systems on the nanoscale have been shown to exist, 
macroscopic quantum effects \cite{Peano04} should be readily 
observable. A second class of experimental systems addresses the resonant non-destructive read-out of a
persistent current qubit \cite{Lupascu03,Bertet04,Lee05,Siddiqi05}. In contrast to the conventional switching
current measurement that generates unwanted quasi-particles when the dc-SQUID (acting as the qubit detector)
switches to the voltage state, this technique keeps the SQUID biased along the supercurrent branch during the
measurement. Thereby, the Josephson plasma resonance of the SQUID depends on the inductive coupling of the
SQUID to the qubit. Measuring the plasma resonance allows to non-destructively read out the qubit state.  
The application of this read-out technique in the nonlinear regime of the SQUID could allow for an improved
sensitivity as well as its potential use as a nonlinear amplifier. 
Finally, we note that there exists a wide parameter regime (typically strong driving and/or big
nonlinearities) where a quantum chaotic behavior of the system with many interesting 
features can occur \cite{Pechukas82,Pechukas84}. A detailed study of this regime goes beyond the scope of this
work. 

To summarize, we have investigated the nonlinear response of the amplitude as well as of the phase of the
oscillations of a driven damped anharmonic resonator (quantum Duffing oscillator). The use of an efficient 
Floquet-Born-Markovian master equation allows to determine the stationary long-time solution directly via
straightforward diagonalization of the rate matrix. This allows to study the amplitude as well as the phase
response for a wide range of parameters. Most importantly, we find pronounced resonances as well as
antiresonances which are associated to multi-photon transitions in the resonator. We have found an interesting
non-monotonous behavior of the antiresonance including a cross-over to a true resonant peak. This cross-over
can  be enhanced by the presence of the bath but is already inherent in the underlying coherent driven
quantum system.  Furthermore, we have found a clear separation of time scales in the dynamics {\em how\/} the
globally stable state is approached starting from the metastable state. This tunneling process can be
characterized by a single tunneling rate $\Gamma$ which follows straightforwardly as the smallest non-zero
eigenvalue of the rate matrix. If the control parameter being the frequency $\omega$ is varied, resonant
tunneling between the two states  clearly is discerned as peaks in the tunneling rate. We hope that these rich
features of the quantum Duffing oscillator will be found in future experiments. 
\section{Acknowledgments}
We are grateful to H.\ Postma for interesting discussions on suspended carbon nanotubes. 
This work has been supported by the German DFG SFB/TR 12.    

\vspace*{40mm}

\begin{figure}[t]
\begin{center}
\epsfig{figure=fig6.eps,width=80mm,keepaspectratio=true}
\caption{(Color online) Five-and six-photon resonance of the response curve $A(\omega)$ 
for different values of the temperature   $T$. Inset: Local extremum of the
$N-$photon resonance for varying temperature. 
The six-photon resonance develops from an antiresonance for low $T$ 
 to a true resonance for higher $T$. 
Remaining parameters are $f=0.1 f_0, 
 \alpha=0.1\alpha_0, \gamma = 0.005 \omega_0$. 
  \label{fig.adamp}\vspace*{20mm}}
\end{center}
\end{figure}
\begin{figure}[t]
\begin{center}
\epsfig{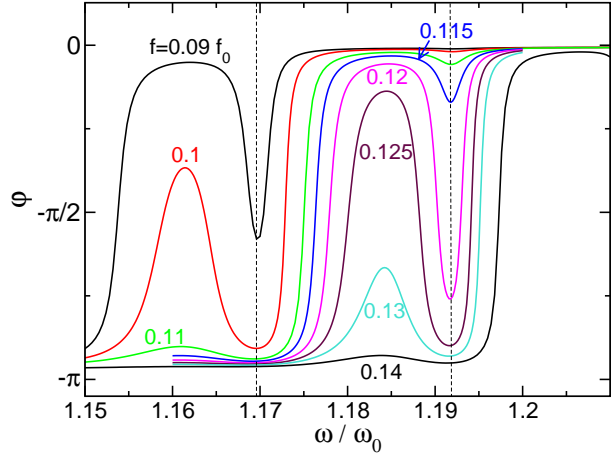}
\caption{(Color online) Phase shift $\varphi (\omega)$ for the 
five-and six-photon resonance 
for different driving amplitudes $f$. 
The dashed vertical lines mark the $N=5-$ and the $N=6-$photon transition.  
Remaining parameters are $T=0.1 T_0, 
 \alpha=0.1\alpha_0, \gamma = 0.005 \omega_0$. 
  \label{fig.phidriv}\vspace*{20mm}}
\end{center}
\end{figure}
\begin{figure}[t]
\begin{center}
\epsfig{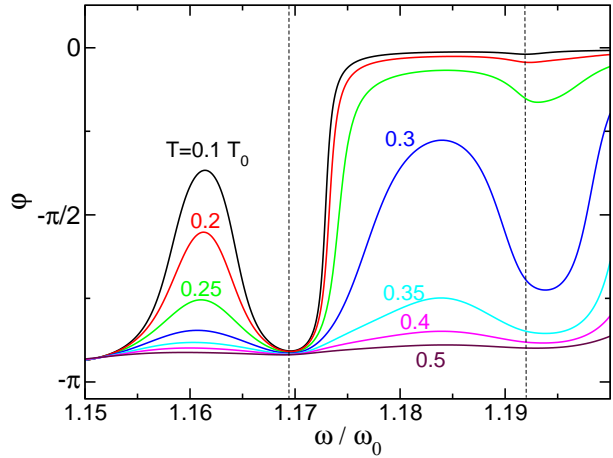}
\caption{(Color online) Phase shift $\varphi (\omega)$ 
for different values of the temperature $T$. 
The dashed vertical lines mark the $N=5-$ and the $N=6-$photon transition.  
Remaining parameters are $f=0.1 f_0, 
 \alpha=0.1\alpha_0, \gamma = 0.005 \omega_0$. 
  \label{fig.phitemp}\vspace*{20mm}}
\end{center}
\end{figure}
\begin{figure}[t]
\begin{center}
\epsfig{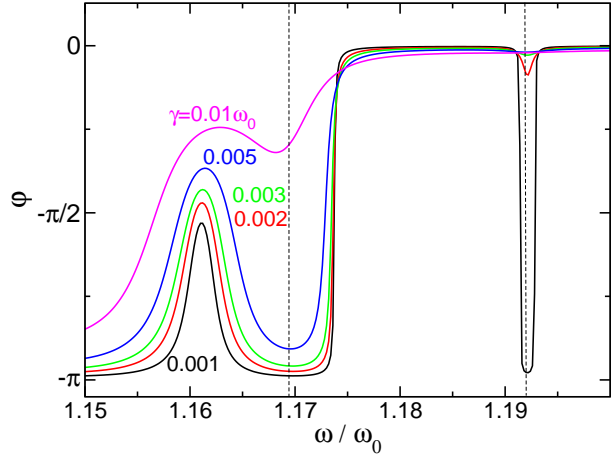}
\caption{(Color online) Phase shift $\varphi (\omega)$ 
for different values of the damping constant $\gamma$. 
The dashed vertical lines mark the $N=5-$ and the $N=6-$photon transition.  
Remaining parameters are $f=0.1 f_0, 
 \alpha=0.1\alpha_0, T = 0.1 T_0$. 
  \label{fig.phidamp}\vspace*{20mm}}
\end{center}
\end{figure}
\begin{figure}[t]
\begin{center}
\epsfig{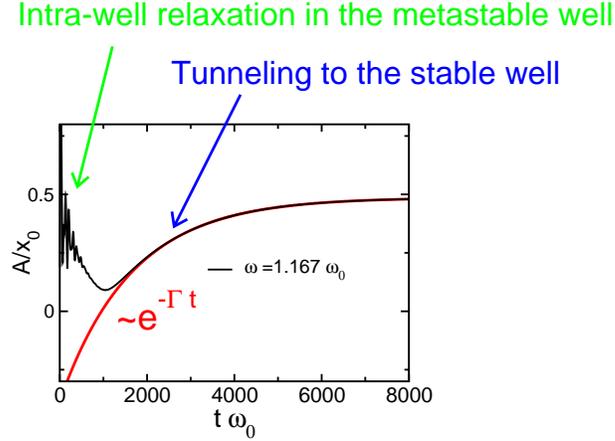}
\caption{(Color online) Time-resolved dynamics of $A$ for $\omega=1.167 \omega_0$ 
(black solid line). Fast transient oscillations occur as ``intrawell'' relaxation in the 
metastable well. The long-time dynamics is governed by a slow exponential decay 
to the globally stable state characterized by a rate $\Gamma$ (tunneling rate). 
The red solid line shows a fit to an exponential 
$~e^{-\Gamma t}$. Here, 
$\alpha=0.1\alpha_0, f=0.1 f_0, k_B T=0.1 \hbar \omega_0$  
and $\gamma=0.005 \omega_0$. \label{fig.timeresolv}}
\end{center}
\end{figure}
\begin{figure}[t]
\begin{center}
\epsfig{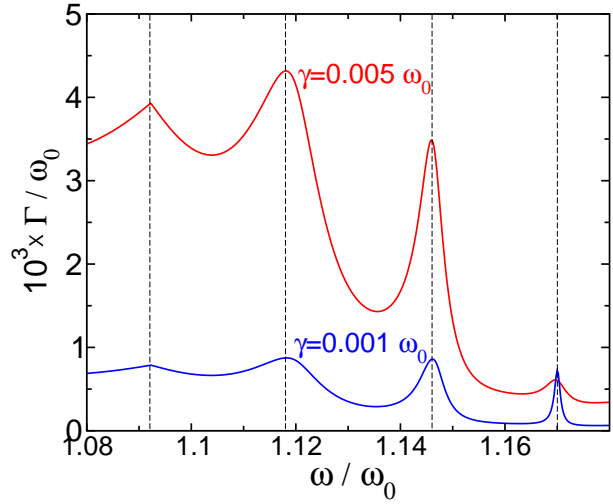}
\caption{(Color online) Tunneling rate  $\Gamma$ for the slow dynamics of the 
amplitude $A$ for approaching the steady state for two different damping strengths. 
The peaks correspond to resonant tunneling in the dynamic bistability. Note that 
tunneling is increased for increasing $\gamma$ (bath-assisted resonant tunneling). 
Moreover, 
$\alpha=0.1\alpha_0, f=0.1 f_0$ and $k_B T=0.1 \hbar \omega_0$. \label{fig.restun}}
\end{center}
\end{figure}

\end{document}